\documentstyle{article}
\newcommand{\ket}[1]{| #1 \rangle}
\begin{document}

\title{Singularities in the Bethe solution of the XXX and
XXZ Heisenberg spin chains}
\author{Rahul Siddharthan\footnote{email:
 \tt rsidd@physics.iisc.ernet.in} \\ 
  \it Department of Physics, Indian Institute of Science, \\ 
  \it Bangalore 560012, India}
\date{Version April 20, 1998; printed \today}
\maketitle
\begin{abstract}
We examine the question of whether Bethe's ansatz reproduces all
states in the periodic Heisenberg XXZ and XXX spin chains. As was
known to Bethe himself, there are states for which the Bethe momenta
$k_n$ diverge: these are in fact the simplest examples of ``string''
solutions. The coefficients of the Bethe wavefunction, too, diverge.
When there are only two down spins in the system (the case considered
by Bethe), we can renormalize these coefficients to get a sensible
(and correct) wavefunction. We show that this is not always possible when
there are more than two down spins. The Bethe equations have several
such divergent solutions, and some of these correspond to genuine
eigenfunctions of the Hamiltonian, but several do not. Nor do they
reproduce the correct energy eigenvalues. Moreover, we point out that
the algebraic Bethe ansatz, an alternative way to construct the
wavefunctions proposed by Faddeev, Takhtajan et al., leads to
vanishing wavefunctions for all these solutions.  Thus, the Bethe
ansatz solution of the Heisenberg model must be regarded as either
incomplete, or inaccurate.
\end{abstract}

\section{Introduction}
The XXZ spin-half Heisenberg Hamiltonian for $N$ interacting
spins, on a one-dimensional chain with periodic boundary
conditions ($n+N \equiv n$), is
\begin{equation} 
H = -2\sum_{n=1}^N \left( S_n^x S_{n+1}^x + S_n^y S_{n+1}^y 
      + \Delta S_n^z S_{n+1}^z \right)
\end{equation}
where $\Delta$ is the anisotropy parameter, $0$ for the XY
model, $1$ for the isotropic ferromagnet. $\Delta=-1$
can be mapped to the isotropic antiferromagnet. We have put the
overall factor of 2 for convenience. In terms of the Pauli
spin matrices,
\begin{equation}
H = -\sum_{n=1}^N \left( \sigma_n^+ \sigma_{n+1}^- + \sigma_n^- 
   \sigma_{n+1}^+ + \frac{\Delta}{2} \sigma_n^z \sigma_{n+1}^z \right)
\label{heis-hamil}
\end{equation}
where $\sigma^\pm \equiv (\sigma^x \pm i \sigma^y)/2$.  This
was among the first many-body quantum problems to be solved
exactly. Bethe \cite{bethe} solved the isotropic model (the ``XXX
model'', $\Delta=1$) in 1931, and the XXZ Hamiltonian can be
solved similarly with little additional effort. The form of
wavefunction which he used has become known as the Bethe
Ansatz and has been applied to a wide variety of other
problems.

The Bethe ansatz, however, is a guess, and while one can
generally verify whether or not it works, there is no easy
way to tell whether it is ``complete'', that is, whether
{\em all} the energy eigenstates are of this form.

There are actually two versions of the Bethe ansatz---the original
method of Bethe, and the ``algebraic Bethe ansatz'' developed by
Takhtajan, Faddeev and others \cite{takhxxz,takhpanch} based on the
work of Baxter \cite{baxter} on vertex models. Though they start from
very different ideas, the two approaches lead to the same equations
and (apart from normalization) the same wavefunctions. Some concrete
statements on completeness have been made in the case of the XXX
model, namely that the algebraic Bethe ansatz reproduces only the
states of ``highest weight'' with respect to the underlying SU(2)
symmetry (which has been proved) and that it reproduces all such
states (which has not been shown strictly). By ``highest weight'' is
meant a state with the maximum allowed value of $S^z$ (the total $z$
component of the spin) for given total spin $S$; such a state is
annihilated by the total spin raising operator $S^+$. The proof that
all Bethe states are of highest weight \cite{takhpanch}, however,
depends on a change of variables from Bethe's $k_n$ to new variables
$\lambda_n$ which arise more naturally in the algebraic method
(besides being convenient in other ways), and the subsequent
restriction that these variables must be finite-valued and distinct.
The original ansatz of Bethe, in terms of the $k$'s, reproduces many
non-highest-weight states without trouble.  Moreover, this doesn't
answer the question of completeness in the anisotropic (XXZ) case
where there is no underlying SU(2) symmetry.  The arguments for
completeness even in the XXX chain are not rigorous: Essler {\em et
al.} \cite{essler}, for instance, show that the usual state-counting
argument \cite{spinwave,takhpanch} is not quite correct, but they view
this as a reorganization of states and suggest that the ansatz for the
XXX chain is SU(2) complete anyway.

Here, after a quick review of the Bethe ansatz, coordinate
and algebraic, mainly to fix the notation, we look into
these questions.  We find that a number of states
correspond to Bethe $k$'s which diverge to $\pm i\infty$,
and these include states which are of highest weight at the
isotropic point. The variables $\lambda_n$ which one
usually uses in the algebraic Bethe ansatz are well-behaved,
but the coefficients of the Bethe wavefunction are singular.
The algebraic method offers an alternative way to produce
a wavefunction, but this construction produces a vanishing
wavefunction.

To handle this situation, one can consider only the
ratios of the Bethe coefficients, rather than the coefficients
themselves (which is reasonable since the wavefunctions
are not normalized). 
These turn out to be finite, but we then find
that some of the states thus produced are not eigenstates
of the Hamiltonian at all. In fact, the energies predicted
by the Bethe equations also turn out to be wrong.

In short, the equations of Bethe, written in terms of the
variables $\lambda_n$, have a number of solutions with
finite, well-behaved $\lambda_n$  which (depending on one's
viewpoint) either do not reproduce the corresponding
eigenstates, or produce too many states including some which
are not eigenstates.

Finally, we also point out that one can formally get rid of
these singularities by introducing an Aharonov-Bohm flux
through the ring, that is, associating different phases with
the forward and backward spin hopping amplitudes. 
But for certain special values of the phase
(namely, the $N$ roots of unity---corresponding to a total flux
of a multiple of $2\pi$ through the ring) the singularities reappear,
and for values of the flux close to zero or to any of these
points the Bethe equations become numerically ill-behaved
and any rootfinder would have severe problems converging to
a solution.

\section{The coordinate Bethe ansatz}

Since the Hamiltonian (\ref{heis-hamil}) commutes both with
the total spin and with $\sigma_z \equiv \sum_n \sigma_n^z$,
we can work with an Ising basis with fixed $\sigma^z$.
Suppose the number of down-spins is $l$ ($\le N/2$: the
other states can be reached by symmetry), and for a given
Ising state, suppose the positions of the down-spins
are $x_1$, $x_2$, \ldots, $x_l$, and call this basis
state $|x_1 x_2 \ldots x_l \rangle $. A general wavefunction
can then be written in this basis as
\begin{equation}
|\Psi \rangle = \sum_{x_1 < x_2 \ldots < x_l} \psi(x_1,x_2,\ldots,x_l)
                | x_1,x_2,\ldots,x_l \rangle.
\end{equation}
Bethe suggested the following form for the expansion coefficients,
\begin{equation}
\psi(x_1,x_2,\ldots,x_l) = \sum_P A(P) \exp \left( i\sum_{n=1}^l k_{P_n}x_n
                              \right)
\label{bethewf}
\end{equation}
where $P$ is a permutation of the integers $1$, $2$, \ldots, $l$, 
the sum is over all permutations with amplitudes $A(P)$, and the 
$k$'s are some as yet undetermined quantities.

One can show that such a wavefunction solves the Hamiltonian 
(\ref{heis-hamil}) if the amplitude for exchange of two
neighbouring particles is
\begin{equation}
A_{mn} \equiv \frac{A(\ldots,m,n,\ldots)}
               {A(\ldots,n,m,\ldots)}
      = -\frac{e^{i(k_m +k_n)} - 2\Delta e^{ik_m} + 1}
             {e^{i(k_m+ k_n)} - 2\Delta e^{ik_n} + 1} .
\end{equation}
In addition,  periodic boundary conditions imply that
\begin{equation}
e^{ik_m N} = \prod_{n \neq m} A_{mn} = 
   (-1)^{l-1} \prod_{n \neq m} \frac{e^{i(k_m+ k_n)} - 2\Delta e^{ik_m} + 1}
             {e^{i(k_m +k_n)} - 2\Delta e^{ik_n} + 1} .
\label{keqns}
\end{equation}
Equations (\ref{keqns}) can be solved for $k_n$, and each solution
\{$k_n$\} gives us an eigenstate of (\ref{heis-hamil}), with
associated energy
\begin{equation}
\label{kenergy}
E = -N \Delta /2 + 2 l \Delta -2 \sum_{n=1}^l \cos (k_n).
\end{equation}

For notational simplicity, we will often use
\begin{equation}
z_n = e^{i k_n}
\end{equation}
so that the Bethe equations become
\begin{equation}
\label{zeqns}
{z_m}^N = (-1)^{l-1}  \prod_{n \neq m} \frac{z_m z_n -2 \Delta z_m + 1}
              {z_m z_n - 2 \Delta z_n + 1}.
\end{equation}

\section{The Algebraic Bethe Ansatz} \label{ABA}

The same equations can be derived rather differently, by diagonalizing
the transfer matrix of the ``six vertex model'', in which the Heisenberg
XXZ Hamiltonian is embedded. Here we will concern ourselves only with the
XXX model, where the algebra is much simpler. We will not describe this
method (see \cite{takhpanch}, for instance) but only mention the change
of variables involved, and how the eigenvectors are constructed.

In the case of the XXX model, we define new variables $\lambda_n$
as follows:
\begin{eqnarray}
z_n &= &\frac{\lambda_n +i/2}{\lambda_n-i/2}  \\
\lambda_n & = & \frac{i}{2} \frac{z_n + 1}{z_n - 1} 
           ~~=~~ \frac{1}{2} \cot (k_n/2).
\end{eqnarray}
These variables arise very naturally in the algebraic approach.
In terms of these, the Bethe equations (\ref{keqns}) become
\begin{equation}
\left(\frac{\lambda_m + i/2}{\lambda_m -i/2} \right)^N
  = \prod_{n \neq m} \frac{\lambda_m - \lambda_n + i}
     {\lambda_m-\lambda_n-i}, \label{leqns}
\end{equation}
and the energy of the corresponding state is
\begin{equation}
E = -\frac{N}{2} + \sum_n \frac{4}{1+4 {\lambda_n}^2}.
\label{lenergy}
\end{equation}

The commuting family of transfer matrices for the six vertex
model, $T(\lambda)$, embed the Heisenberg Hamiltonian,
and the eigenvectors of the former are the eigenvectors of
the latter.  $T(\lambda)$ is the trace of the monodromy matrix 
\begin{equation}
\tau(\lambda) = L_N(\lambda) L_{N-1}(\lambda)\ldots L_1(\lambda)
\end{equation}
where $L_n(\lambda)$ is the ``local $L$ operator''
\begin{equation}
L_n(\lambda) = \left( \begin{array}{cc}
                       \lambda+(i/2)\sigma_n^z & i \sigma_n^- \\
                       i\sigma_n^+ & \lambda-(i/2)\sigma_n^z
                      \end{array} \right).
\end{equation}
Writing the monodromy matrix as
\begin{equation}
\tau(\lambda) = \left( \begin{array}{cc}
                       A(\lambda) & B(\lambda) \\
                       C(\lambda) & D(\lambda) 
                       \end{array} \right),
\end{equation}
it can be shown that the vector
\begin{equation}
\ket{\{\lambda_n\}} = \left( \prod_{n=1}^l B(\lambda_n) \right) 
         \ket{\uparrow\uparrow\uparrow\cdots}
\end{equation}
is an eigenvector of the transfer matrix $T$, and therefore
of $H$, if the $\{\lambda_n\}$ satisfy the equations (\ref{leqns}).

\section{Completeness}
\label{complete}

Since the Hamiltonian (\ref{heis-hamil}) is translationally invariant, a
useful classification of the energy eigenstates is in terms of eigenstates
of the translation operator $T$ which shifts all spins by one lattice
site to the left. All eigenstates of $H$ can be chosen to be
eigenstates of $T$ (non-degenerate eigenstates are necessarily so
anyway) with eigenvalues $\phi$ such that $\phi^N=1$.
The Bethe ansatz does this automatically: for given \{$z_n$\},
$\phi = \prod z_n$.

There is a limit where the Bethe method does reproduce all states
without trouble: that of $\Delta=0$, or the XY model, which
is equivalent to a free Fermi system. The exchange amplitudes
$A_{mn}$ in (\ref{keqns}) each become $-1$, and
the $k$'s are $2\pi i/N$ times an integer (for odd $l$) or
a half-integer (for even $l$), all distinct (otherwise
the wavefunction vanishes). We then have exactly as many states as we
need. Some of these states evolve
smoothly as one turns on $\Delta$, but some don't,
and it's interesting to see why not.

It turns out that the $\Delta=0$ states are sometimes degenerate, but
the degeneracy is lifted for any finite $\Delta$. The
states, if chosen correctly, do evolve continuously as
$\Delta$ is changed from zero, but the $\Delta=0$ Bethe ansatz
gives superpositions of these which are no longer eigenstates
for any finite $\Delta$. We see concrete examples of this below.
This also suggests a way around the problem: introduce an
Aharonov-Bohm flux, which does not change the eigenstates
at $\Delta=0$ but lifts the degeneracies in energy. We
discuss this later.

\subsection{Four sites}
\label{foursite}

The generic Bethe Ansatz wavefunction for two down-spins is
\begin{equation}
\ket{\psi} = \sum_{x_1,x_2} \left( {z_1}^{x_1} {z_2}^{x_2} + 
                    \frac{1}{A} {z_2}^{x_1}{z_1}^{x_2} \right) \ket{x_1,x_2}
\label{generic4}
\end{equation}
with $x_1 < x_2$; and we also have
\begin{eqnarray}
{z_1}^4 &=& A, \label{zAcond} \\
 {z_2}^4 &=& 1/A, \\
 A &=& - \frac{z_1 z_2 -2 \Delta z_1 +1}{z_1 z_2 -2\Delta z_2 + 1}.
\end{eqnarray}
We will number sites starting from $0$, to ease the algebra.

Let us follow the convention of denoting a down-spin by $1$ and
an up-spin by $0$. In the case of four sites and two down-spins,
we can write down the eigenstates of $T$, with corresponding
eigenvalues $\phi$, immediately:

\begin{eqnarray}
\phi=1: ~~~
\ket{\psi_1} &=& \ket{1100}+\ket{0110} + \ket{0011} + \ket{1001} \nonumber \\
\ket{\psi_2} &=& \ket{1010}+ \ket{0101} \nonumber \\
\phi=i: ~~~
\ket{\psi_3} &=& \ket{1100}+i\ket{0110} - \ket{0011} -i\ket{1001}\nonumber \\
\phi=-1: ~~~
\ket{\psi_4} &=& \ket{1100} - \ket{0110} + \ket{0011} - \ket{1001}\nonumber \\
\ket{\psi_5} &=& \ket{1010} - \ket{0101} \nonumber  \\
\phi=-i: ~~~
\ket{\psi_6} &=& \ket{1100} - i\ket{0110} - \ket{0011} +i\ket{1001} \nonumber
\end{eqnarray}

Now, for $\phi=0$ one has to further diagonalize $H$ in the subspace of
$\ket{\psi_1}$ and $\ket{\psi_2}$; but it turns out that all the rest
are already eigenstates of $H$. Moreover, $\psi_4$ and $\psi_5$ at
$\Delta=0$ are degenerate, and the Bethe ansatz gives us some superposition
of these; but the degeneracy is lifted for arbitrarily small finite
$\Delta$. This is one reason the completeness question is nontrivial:
it is not obvious that new well-behaved solutions of the Bethe
equations will emerge for finite $\Delta$, and if they do, they must
be ill-behaved as $\Delta \rightarrow 0$ 
since we already have all the solutions there.

In fact, $\ket{\psi_5}$ is reproduced with the choice ($z_1$,$z_2$) = 
$\pm 1$. This solution works for all $\Delta$ except 0. We now
demonstrate that there is no well-behaved solution for $\ket{\psi_4}$.

From the generic Bethe wavefunction in this case (\ref{generic4}),
we see that $z_1 z_2 = -1$. On the other hand, the coefficient of
$\ket{1010}$ should be zero, so we have ${z_2}^2 + (1/A){z_1}^2 =0$.
Combining these conditions gives us ${z_1}^4 = -A$ which contradicts
(\ref{zAcond}) unless $z_1=0$, which implies $A=0,  z_2=\infty$.

All the other states evolve smoothly as $\Delta$ is turned up from
zero; however, singularities occur at particular values for $\Delta$.
At $\Delta=1$, the upper of the two $\phi=0$ states corresponds to
$z_1 = z_2 = 1$ so the Bethe wavefunction is singular (or zero).
The same thing happens to the $\phi = \pm i$ states at $\Delta = \sqrt 2$.
Generally this ``collision'' of $z$'s or $k$'s occurs at the point
where the $k$'s go complex. One can make sense of it,
however, by considering the limit of the wavefunction as
$\Delta$ approaches the critical value.

In terms of the algebraic Bethe ansatz (we confine ourselves
here to the XXX point), things are a little different: if we
require that the $\lambda$'s, rather than the $k$'s or the
$z$'s, be finite, the value $z=1$ is forbidden. Then the
states $\ket{\psi_3}$, $\ket{\psi_5}$ and $\ket{\psi_6}$
(which are not highest weight states) are not reproduced.
It may appear that the situation is saved for the
highest-weight state $\ket{\psi_4}$: the values $0$ and
$\infty$ for $z$ translate to the nicer values $\pm i/2$ for
$\lambda$.  The energy (\ref{lenergy}) appears to be
divergent, but if we take the $\lambda$'s to be
$(-i/2+\epsilon, i/2+\epsilon)$ and let $\epsilon$ vanish,
we do get the correct value.  Unfortunately, these $\lambda$
values don't produce an eigenstate:
$B(i/2)B(-i/2)\ket{\uparrow\uparrow\uparrow\uparrow} =0$,
which is easy to verify.

So there exists at least one state, $\ket{\psi_4}$, which is
not reproduced by the algebraic Bethe ansatz at the
isotropic point.   In fact a similar singularity exists with
the state 
\[ 
\sum (-1)^n \sigma_n^- \sigma_{n+1}^- \ket{\uparrow\uparrow\uparrow\cdots},
\]
for all even $N$. This state was actually known to Bethe \cite{bethe},
who remarks that the $k$'s must diverge to accommodate it.

\subsection{Six sites}
\label{sixsite}
Here things get a little more complicated and we have many more
states to contend with. But classification of
states in terms of $\phi$, the eigenvalue of the translation
operator $T$, can still be done. In case of a real translation
eigenvalue $\phi$ ($\pm 1$) we can further consider the parity
operator, which reverses the order of the spins. (If $\phi$
is not real, this doesn't commute with $T$, but one can generalize
the definition to take care of that.)  So consider
the state
\begin{eqnarray}
\ket{\psi} = & & \ket{110100} + \ket{011010} + \ket{001101}+\ldots \nonumber \\
            & & -\ket{101100} - \ket{010110} -\ldots
\label{anom6_1}
\end{eqnarray}
which is the only state with $\phi=1$ and parity $-1$, and is
therefore an eigenstate of $H$ for arbitrary $\Delta$
(also, as one can verify, a highest weight state).
Consider also the state 
\begin{eqnarray}
\ket{\psi'} = & & \ket{110100} - \ket{011010} + \ket{001101}+\ldots\nonumber\\
            & & +\ket{101100} - \ket{010110} + \ket{001011} -\ldots
\label{anom6_2}
\end{eqnarray}
which is also an eigenstate of $H$ (the only one with $\phi=-1$
and parity 1), and a lowering of a state analogous to $\ket{\psi_4}$
in the previous section. Neither of these states can be
reproduced unless either two of the $z$'s are equal, or
one of them vanishes (and, therefore, another diverges). Since
the proof is not as compact as in the four site case, 
we leave it for appendix \ref{appA}. These states, too, are degenerate with
other states at the XY point, and some superpositions of these are
reproduced by the Bethe ansatz.

In appendix \ref{appC}, we present the values of $\lambda$
which produce these states at the isotropic point if we take
appropriate limits for the divergent Bethe coefficients. As
predicted, the corresponding values of $k$ are divergent.

\subsection{Adding a flux}

The XY degeneracies can be lifted by passing an Aharonov-Bohm
flux through the ring. This modifies the forward and backward
hopping amplitudes by phase factors:
\begin{equation}
H = \sum_{n=1}^N \left( \alpha^* \sigma_n^+ \sigma_{n+1}^- 
                    + \alpha \sigma_n^- \sigma_{n+1}^+
       + \frac{\Delta}{2} \sigma_n^z \sigma_{n+1}^z \right),
\end{equation}
where $\alpha$ is a complex number of unit modulus. For $\alpha=1$
we recover the original Hamiltonian. For most other values
of $\alpha$, the ground state degeneracies are lifted and the
states evolve smoothly when one turns on $\Delta$. But
the singularities reappear when one removes the flux. Moreover,
certain other values of $\alpha$ also prove
to be singular. This is an example of the fact, noted by Byers and
Yang \cite{byers}, that the energy levels of a quantum system on
a ring are periodic in the flux through the ring. 
To see it in this case, we carry out a Bethe ansatz solution
and get the analogue of (\ref{zeqns}):
\begin{equation}
{z_m}^N = (-1)^{l-1} \prod_{n \neq m} 
              \frac{\alpha^* z_m z_n -2\Delta z_m + \alpha}
               {\alpha^* z_m z_n -2\Delta z_n + \alpha},
\end{equation}
which, if we define $w_n = \alpha z_n$, becomes
\begin{equation}
{w_m}^N \alpha^N = (-1)^{l-1} \prod_{n \neq m} 
              \frac{w_m w_n -2\Delta w_m + 1}
               { w_m w_n -2\Delta w_n + 1}.
\end{equation}
These are identical to the Bethe equations if $\alpha^N =1$, so
for these values of $\alpha$ the singularities will persist.

The exact solution of the $\phi=-1$ sector for four sites with flux is
given in appendix \ref{appB}, and its behaviour as the
flux is made to vanish can be seen explicitly there.

\subsection{Longer chains}

The common thing to the above examples is the divergence of one of the
$z$'s, or of the
$k$'s to $\pm i \infty$.  In fact for any system size and any number
of overturned spins, if one of the $z$'s diverges at
$\Delta=1$, the
algebraic Bethe ansatz fails to reproduce the corresponding
state. To see this in the XXX case, 
note that if one $z$ diverges, another must
vanish; the $\lambda$'s corresponding to these are $\pm i/2$.
Consider the general algebraic Bethe state:
\[ \ket{\{\lambda_n\} } = \prod B(\lambda_n) \ket{\uparrow \uparrow
                                   \uparrow \cdots} 
\]
Choosing the argument of the last $B$ operator to be $i/2$,
we can easily convince ourselves (from the definition of $B$
in terms of the $L$ matrices in sec.\ \ref{ABA}) that the 
action of this on the
ferromagnetic state is merely to flip the first spin. If we
choose the next $B$ operator to have the argument $-i/2$ that
state is killed entirely. The values of the other $\lambda$
variables then don't matter.

It is also quite clear that such solutions will in general
exist for all lattice lengths. To see this, plug the values
$\lambda_l = i/2$, $\lambda_{l-1}=-i/2$ into eq. (\ref{leqns}).
These equations are immediately satisfied for $m=l$ and
$m=l-1$. For the rest, they now become
\begin{equation}
\left(\frac{\lambda_m + i/2}{\lambda_m -i/2} \right)^{N-1}
\frac{\lambda_m - 3i/2}{\lambda_m + 3i/2}
  = \prod_{{n=1} \atop {n \neq m}}^{l-2} \frac{\lambda_m - \lambda_n + i}
     {\lambda_m-\lambda_n-i}, \label{leqns_singular}
\end{equation}
and any solution of these for $\lambda_1$, \ldots, $\lambda_{l-2}$
combined with the above values for $\lambda_{l-1}$, $\lambda_l$ is
a solution of (\ref{leqns}).

This, by the way, is the simplest sort of ``string''
solution---such solutions, of groups of complex $\lambda_n$
with a common real part and the imaginary parts separated by
$i$, arranged symmetrically about the real axis, 
are generic in the thermodynamic
limit. This particular string exists for all system sizes.

Therefore for any lattice length $N$ and spin sector $l$
there exist several solutions for the equations
(\ref{leqns}) for which the algebraic Bethe ansatz gives a
vanishing wavefunction and the coordinate Bethe ansatz a
singular one. How to handle these singularities, and whether
they can give us meaningful wavefunctions,
are discussed later.

Though we have discussed only the XXX model above, the same
is true of the XXZ model. The variables $\lambda_n$ are
defined a little differently here (the earlier ones are
related to these by a scale factor) and the Bethe equations
are \cite{takhpanch}
\begin{equation}
\left(\frac{\sin(\lambda_m+\eta)}{\sin(\lambda_m-\eta)} \right)^N
= \prod_{m \neq n} \frac{\sin(\lambda_m-\lambda_n+2\eta)}
              {\sin(\lambda_m-\lambda_n-2\eta)}.
\end{equation}
Here $\Delta=\cos 2\eta$.
These have the singular solution $\lambda_{l-1},\lambda_l =\pm \eta$
with $l-2$ equations for the remaining $\lambda_m$. The $L$
matrix in this case is
\begin{equation}
L_n(\lambda) = \left( \begin{array}{cc}
  w_4(\lambda) + w_3(\lambda) {\sigma_n}^z & (\sin 2\eta) {\sigma_n}^- \\
  (\sin 2\eta) {\sigma_n}^+ & w_4(\lambda)-w_3(\lambda){\sigma_n}^z
  \end{array} \right)
\end{equation}
with $(w_4 + w_3)(\lambda) = \sin(\lambda+\eta)$ and $(w_4-w_3)(\lambda)$
$=$ $\sin(\lambda-\eta)$.
The $B$ operator is defined as before, and one can convince oneself
that this choice of $\lambda_n$ annihilates the reference state, just
as in the XXX case.

Unfortunately our proofs regarding the specific singular states
are not readily generalized to longer chains, so other than
the one example at the end of sec.\ \ref{foursite}, and the states
suggested at the end of appendix \ref{appC}, we have no
rigorous examples to give. But the above examples involve states
with $\delta$-independent coefficients, which are degenerate
with other states at $\delta=0$, and such states certainly
exist for larger lattices as well, so it is entirely conceivable 
that they too could pose problems for the Bethe ansatz.

We can be a little more specific. In the XY limit, the $k$'s are 
the ``bare'' momenta $2\pi I_n/N$, where $I_n$ are distinct
integers (for odd $l$) or half-integers (for even $l$). Consider
the sector $\phi=1$ ($\phi$ being, as above, the eigenvalue of $T$);
this is the same as saying $\sum k_n =0$. This can happen in
two ways: either the $k_n$ are symmetrically distributed around $0$,
or they are not. If they are not, the solution corresponding to
$-k_n$ is distinct from this solution but degenerate in energy with
it. However, these two solutions, generally, do not evolve into
well-behaved solutions as one turns on $\Delta$: the solutions at
finite $\Delta$ correspond to superpositions of these, which are no
longer degenerate. The same problem occurs in
the $\sum k_n = \pi$ sector.

Numerically, we investigated several such states. The idea was to
solve equations (\ref{zeqns}) using a Newton-Raphson rootfinder
(capable of finding complex roots: a simple modification of the
standard method). For this method to
work, one needs a good starting guess. So a possible method is
to keep a finite flux $\Phi$, start the rootfinder from the XY point
where the solution is known (and not affected, except in energy,
by the flux), evolve $\Delta$ forward to some desired value
and then evolve the flux down to zero. In every case, the last step
failed to converge, yielding divergent $z$'s or $k$'s or else
hitting some spurious solution which has two $k$'s equal and
therefore yields a vanishing wavefunction.

It seems therefore that such singularities are very common,
at least in the $\phi=1$ and $\phi=-1$ sectors, possibly in
the other sectors too.

\section{Handling the singularities}

Thus, for several eigenstates of the Heisenberg model
the Bethe ansatz becomes ill-behaved and the coefficients
of the wavefunction, diverge (in the coordinate Bethe ansatz)
or vanish (in the algebraic method). For the coordinate
method, one can handle the divergence
by considering only the ratios of the coefficients, rather than
the coefficients themselves. For the algebraic method, since
the problem values of $\lambda$ are $\pm i/2$, one can let
them approach these values and consider the limiting ratios
of the coefficients of the wavefunction. While this should
work in principle, it would be rather messy and we don't
pursue it further: the algebraic BA is valuable more as a
formal elegant unifying tool than as a practical
calculational aid.

For the four site, two spin down case, if we take $\lambda_1=i/2$,
$\lambda_2=-i/2$, this works fine and we reproduce the state
$\ket{\psi_4}$ of sec.\ \ref{foursite}. This is
demonstrated in appendix \ref{appB}. However, consider the
six site case with three down spins. If we choose $\lambda_2$
and $\lambda_3$ to be $\pm i/2$, we are left with one Bethe
equation for $\lambda_1$ which turns out to be a fifth degree
equation with six solutions if one includes $\lambda_1 = \infty$
(or $z_1 = 1$). That, and $\lambda_1=0$, do lead to valid
eigenstates which are precisely the problematic eigenstates
discussed in sec.\ \ref{sixsite}. But the other four solutions
for $\lambda_1$, though perfectly valid solutions of the Bethe
equations (\ref{leqns}), do not lead to correct
wavefunctions (nor, indeed, to correct energies). 
This is further discussed in appendix \ref{appC}.
For any even lattice length with three down spins, the Bethe
equations have $N$ such singular solutions of which two yield
genuine eigenstates and the rest are spurious. This presumably
happens for all spin sectors, but we don't have good estimates
on what fraction of the total number of states is so affected.

Another possible way to handle these singularities (when solving
the equations numerically) is to introduce a
finite flux, and slowly turn it down to zero. It turns out that
the equations are so ill-behaved that this does not help. A
rootfinder such as the Newton-Raphson method, which is based on
the Taylor series and requires smoothly varying functions, behaves
very badly in the neighbourhood of such singularities.

Other kinds of singular Bethe states are possible---for instance,
at certain special values of $\Delta$, two $k$ values may collide.
If the corresponding Bethe state has a well-defined limit as $\Delta
\rightarrow \Delta_0$, where $\Delta_0$ is the value where the $k$'s
coincide, then there is no problem. This happens for instance with
the $\phi=\pm i$ and one of the $\phi=1$  states in sec.\ \ref{foursite},
at $\delta=\sqrt{2}$ and $\delta=1$ respectively. And then again,
the $k$'s may be well-behaved but the $\lambda$'s singular, as in
the state $\ket{1010}-\ket{0101}$. Here, if one just works with
the $k$'s, no problems arise. In particular, the Bethe coefficients
are well-behaved.

\section{Conclusion}

The Bethe ansatz is a particular guess for a wavefunction and
substituting it in the Hamiltonian, and applying periodic
boundary conditions, leads to certain conditions on the parameters
of the wavefunction, the $k$'s or the $\lambda$'s as one may
prefer. We find that though Bethe's equations for the $\lambda$'s
by themselves have a large number of solutions,
some of these solutions lead to singular, ill-defined wavefunctions.
These problems arise from divergences in 
Bethe's original quasimomenta $k_n$,
though the $\lambda_n$ to which they can be transformed are well
behaved.

We give explicit examples of this for four and six site lattices,
but show that such solutions of the Bethe equations should occur for
all lattice sizes. The question then is whether to allow such solutions
or not. If we do not allow them, the Bethe ansatz is incomplete.
If we do allow them, we get vanishing wavefunctions with the
algebraic Bethe ansatz, and divergent ones with the
coordinate method. If we attempt to make sense of these by
considering only ratios of the Bethe coefficients, we get
well-defined wavefunctions but not all of these are
eigenstates of the Hamiltonian.  Indeed, even without
calculating the wavefunctions, we can check that the
energies are wrong.

Earlier arguments for completeness have been based on counting of
solutions to the Bethe equations (\ref{leqns}), using the string
hypothesis. But as we have seen, not every solution to the Bethe
equations produces an eigenstate of the Hamiltonian. Therefore,
even if the Bethe equations themselves have as many solutions as
we require, it does not imply completeness of the method.

In the case of two down-spins, however, the Bethe ansatz is
complete if one normalizes the divergent wavefunctions appropriately.
This was known to Bethe \cite{bethe}, 
as was the fact that the $k$'s and the
amplitudes for the wavefunction diverge for one state in this case.
There are no incorrect solutions in this case.

One question to be answered is how many incorrect solutions do
we obtain in the thermodynamic limit. In the case of three down
spins, we get $N-2$ wrong answers, which is a vanishing fraction
of the total number of states in the thermodynamic limit. 
It will probably remain a vanishing fraction if the number of
down spins is finite, but if a finite fraction of the spins
are down---if $S^z=0$, for
instance---we haven't formed an estimate.

One way to get around this, in principle, is to add a small flux
to the system; but this helps only in a formal sense. The equations
are very ill behaved near these singularities and any rootfinder would
have a great deal of trouble converging to a solution.

Finally, some remarks on why all this is important. One reason
is a matter of principle---it is good to know to what extent
a given method solves a problem. Another is that the question
could be relevant to other problems too---for instance, the
Hubbard model, which becomes the Heisenberg antiferromagnet in
one limit, may also exhibit similar singularities in its Bethe ansatz
solution. This incompleteness will probably not
matter much in calculations of thermodynamic
properties, since the singular states may then be a vanishing
fraction of the total number of states.  However, when
considering small systems, or systems with very few overturned spins such
as low-lying excitations above the ferromagnetic state,
perhaps one should keep the completeness question in mind.
And finally, the above considerations are very important indeed
if one wants to use the Bethe ansatz for numerical calculations on
a finite sized lattice.  This should not be a common difficulty
since other methods, such as exact diagonalization, can easily be
used in one dimension.

\section{Acknowledgements}
I am greatly indebted to B.~Sriram Shastry and Diptiman Sen for
detailed and very enlightening discussions. I also want to thank
Ashwin Pande and Alexander Punnoose for useful comments and
suggestions.

\vspace{1cm}
\appendix
\section{Singular states in the six site lattice}
\label{appA}

For six sites and three down-spins, we have three $z$ variables
and three two-particle scattering amplitudes $A_{mn}$ (since
$A_{nm} = 1/A_{nm}$.
Let $A_{123}=1$; then we have
\begin{eqnarray}
A_{132} = \frac{1}{A_{23}},~~~~ A_{213} = \frac{1}{A_{12}},~~~~ 
     A_{231}= \frac{1}{A_{12}A_{13}}, \nonumber \\
A_{312}=\frac{1}{A_{23}A_{13}},~~~~ A_{321}=\frac{1}{A_{23}A_{13}A_{12}}.
\end{eqnarray}
Also, from the Bethe ansatz equations we have
\begin{equation}
{z_1}^6 = A_{12}A_{13},~~~~ {z_2}^6=\frac{A_{23}}{A_{12}}, ~~~~
    {z_3}^6 = \frac{1}{A_{13}A_{23}}
\end{equation}
Let us define $A_{23}=A$ and rewrite $A_{12}, A_{13}$ in terms
of this and the $z$'s.
Using all these in the formula (\ref{bethewf}) for the Bethe
wavefunction, we get for $\psi(x_1,x_2,x_3)$ the formula
\begin{eqnarray}
\psi(x_1,x_2,x_3)&=&{z_1}^{x_1}{z_2}^{x_2}{z_3}^{x_3}
+\frac{1}{A}{z_1}^{x_1}{z_3}^{x_2}{z_2}^{x_3} 
+\frac{1}{A}{z_2}^{x_1 +6}{z_1}^{x_2}{z_3}^{x_3}  \nonumber \\
& &+{z_2}^{x_1}{z_3}^{x_2}{z_1}^{x_3-6} 
+{z_3}^{x_1+6}{z_1}^{x_2}{z_2}^{x_3} 
+\frac{1}{A}{z_3}^{x_1}{z_2}^{x_2}{z_1}^{x_3-6} \label{bethe6}
\end{eqnarray}
Now consider the state (\ref{anom6_1}) in sec.\ \ref{sixsite}.
To reproduce this state, we require that
\begin{eqnarray}
\psi(0,1,2) & = & 0, \label{cond1} \\
\psi(0,2,4) & = & 0. \label{cond2}
\end{eqnarray}
Each of these, together with $z_1 z_2 z_3 = 1$, 
can be used to solve for $A$. Using (\ref{cond2}), we get
\begin{equation}
A = - \frac{{z_2}^2}{{z_3}^2}
\end{equation}
and using (\ref{cond1}), 
\begin{equation}
A = - \frac{{z_1}^3 z_2 + {z_1}^3 {z_2}^4 + z_2}
            {{z_1}^3 z_3 + {z_1}^3 {z_3}^4 + z_3}
\end{equation} 
Unless $z_2$ or $z_3$ is zero, we can equate these,
and after cancelling a common $z_2/z_3$ factor, rearranging
and using $z_1 z_2 z_3 = 1$, we get
\begin{equation}
{z_1}^3(z_2-z_3) -{z_1}^2({z_2}^2-{z_3}^2) + (z_2-z_3) = 0
\end{equation}
and if we multiply the last term by $z_1 z_2 z_3$, this can
be factorized as
\begin{equation}
z_1 (z_2-z_3) (z_3-z_1)(z_1-z_2) = 0.
\end{equation}

Thus, if the variables $z_1$, $z_2$, $z_3$ are all distinct,
one of them must vanish (and consequently another must diverge).

A precisely analogous proof goes through for the state (\ref{anom6_2})
using the condition $z_1 z_2 z_3 = -1$.

\section{Some exact solutions in the four site case}
\label{appB}

Here, as in section \ref{complete}, we work with states
labelled by $\phi$, the eigenvalue of the translation
operator $T$ which shifts all spins by one lattice site
to the left.

\subsection{Sector $\phi=-1$ with flux $\alpha$}

We have two solutions in this sector, given by
\begin{equation}
(z_1,z_2) = \frac{-\gamma \pm \sqrt{\gamma^2+4}}{2}
\end{equation}
where $\gamma$ has the two possible values
\begin{eqnarray}
\gamma &=& \frac{\Lambda \pm \sqrt{\Lambda^2-8}}{2}, \\
\Lambda &=& \frac{2\Delta}{\alpha-\alpha^*}.
\end{eqnarray}

Note that as $\alpha \rightarrow 1$, $\Lambda$ diverges, and
consequently one solution for $z_1,z_2$ diverges, while the
other approaches $\pm 1$.

\subsection{Sector $\phi=-1$ with no flux}
The state $\ket{1010}-\ket{0101}$ is readily reproduced with
$z=\pm 1$. To reproduce the other state, we need $z_1 \rightarrow 0$,
$z_2 \rightarrow \infty$.
Now the Bethe coefficient of the state $\ket{1010}$ is 
${z_2}^2 + {z_1}^2/A$, which on using $z_1 z_2 = -1$ and $z_1^4 = A$
becomes $2{z_2}^2$. The Bethe coefficient of $\ket{1100}$, similarly,
is $z_2 - {z_2}^3$. Both of these diverge as $z_2 \rightarrow \infty$,
but if we take their ratio, this vanishes. (It also vanishes, as it
should, if $z_2 \rightarrow 0$.) So for this state, we can conclude
that the $\ket{1010}$ basis state is of zero weight compared to
$\ket{1100}$, and we do reproduce the correct eigenstate.

\subsection{Sector $\phi=1$ with no flux}
There are two solutions in this sector, given by
\begin{equation}
(z_1,z_2) = \frac{\gamma \pm \sqrt{\gamma^2-4}}{2}
\end{equation}
for each of the two values of $\gamma$
\begin{equation}
\gamma = \frac{\Delta \pm \sqrt{\Delta^2+8}}{2}.
\end{equation}
As $\Delta \rightarrow 1$, one of the $\gamma$ solutions tends to
$2$, for which the $z$ values collide at the value $1$. At this point,
\[ A = -\frac{z_1 z_2 -2 \Delta z_1 +1}{z_1 z_2 -2\Delta z_2 +1}
\]
also becomes ill-defined. If it weren't so, the wavefunction would
vanish. Now, as before, we can consider the coefficients of
$\ket{1010}$ and $\ket{1100}$ and eliminate the ill-behaved $A$
from them, before taking the $\Delta \rightarrow 1$ limit. 
The former, as before, can be written
solely in terms of $z_2$ as $2 {z_2}^2$ and the latter as
$z_2 + {z_2}^3$. Both of these have well defined limits as $z_2
\rightarrow 1$, and the correct wavefunction is reproduced.

\section{Six sites: exact solutions for the singular states}
\label{appC}

Here again we confine ourselves to the XXX model.
Let us look for a solution of the equations for $\lambda_m$, (\ref{leqns}),
for six sites and three down spins, with $\lambda_1=-i/2$ and $\lambda_3
= i/2$. Then we only have to find $\lambda_2$, which we will call
$\lambda$. Equations (\ref{leqns}) are already satisfied for $\lambda_1$
and $\lambda_3$, while for $\lambda_2 = \lambda$ they become
\begin{equation}
\left(\frac{\lambda+i/2}{\lambda-i/2}\right)^6 =
\left( \frac{\lambda + 3i/2}{\lambda-i/2} \right)
\left( \frac{\lambda + i/2}{\lambda-3i/2} \right)
\end{equation}
which have solutions $\lambda=\infty$ (which we will not disallow), 
$\lambda=0$ and the four roots of
\begin{equation}
16\lambda^4 + 40\lambda^2 - 7 = 0.
\end{equation}
For now, we will not specify which root we are considering; for all of
these, the Bethe equations (\ref{leqns}) are satisfied. The $z$ values
corresponding to these are $z_1=0, z_3=\infty$ with various finite
values of $z_2$: $+1$ for $\lambda=\infty$, $-1$ for $\lambda=0$, and
other values for the other four solutions of $\lambda$.

Our Bethe wavefunction (\ref{bethe6}) is then singular because
of the singular values of the $z$'s. Let us, however, assume
\begin{equation}
(z_1,z_2,z_3) = (1/\beta, z, \phi\beta/z) 
\end{equation}
and take the limit of this as $\beta \rightarrow \infty$.
Here $\phi$ is, as before, the eigenvalue of the translation
operator: $\phi^6=1$. In that case, to leading order in $\beta$, the
coefficients of various Ising states in the Bethe wavefunction are
\begin{eqnarray}
\ket{111000} & : & \beta^5 \left(\frac{\phi}{z} + \frac{1}{z^4} \right)
  \nonumber \\
\ket{110100} & : & \beta^5 \frac{1}{z^3} \nonumber \\
\ket{101100} & : & \beta^5 \frac{\phi^2}{z^2} \nonumber \\
\ket{101010} & : & \beta^4 \frac{1+\phi^2+\phi^4}{z^2} \nonumber
\end{eqnarray}
with coefficients of other Ising states being the above multiplied by
appropriate powers of $z_1 z_2 z_3$.
As $\beta \rightarrow \infty$, the coefficient of $\ket{101010}$ becomes
insignificant. First consider the case $z_2=z=\pm 1$, or $\lambda = 0, 
\infty$. If we choose $\phi = \pm 1$ as appropriate, the coefficient
of $\ket{111000}$ can also be made to vanish (to order $\beta^5$) and
we recover precisely the two states discussed in appendix \ref{appA}.

But now consider the other four possible values of $z$. In
all eigenstates, the coefficients of $\ket{101100}$ and
$\ket{110100}$ differ by a factor of $\phi$ (or a power of
it): $\phi$, remember, is a sixth root of unity. Here they
differ by a factor $\phi^4/z$, which is not a power of
$\phi$ since the four values of $z$ other that $\pm 1$ are
not sixth roots of unity. So for these values of $z$, the
states (which are in any case singular) are not eigenstates
of the Hamiltonian even in a limiting sense.

In fact we have checked the parentage of all the exact
eigenstates in the six site case, and apart from the two
discussed above, all are reproduced with finite $z_n$, with
at worst a collision of two values. So there is no room for 
the remaining four solutions. Moreover, if one uses
the values of $\lambda$ calculated in this appendix to
calculate the energy by equation (\ref{lenergy}), using a
limiting method as in sec.\ \ref{foursite} to handle the
divergence for $\lambda_1, \lambda_3=\pm 2$, one gets the
correct answers for $\lambda_2=0,\infty$ but wrong answers
for the other four solutions for $\lambda_2$---as one can
verify with exact diagonalization. (For this calculation,
all we need to assume is that
\begin{equation}
\frac{4}{1+4(i/2)^2} + \frac{4}{1+4(-i/2)^2} =2,
\end{equation}
which as a limit is justified in sec.\ \ref{foursite}, and
which produces the correct energy when the wavefunction is
correct).
So there exist solutions
of the Bethe equations (\ref{leqns}) which do not correspond
to actual eigenstates of the problem.

In the case of three down spins, the above generalizes
quite readily to any even number of sites: for $N$ sites, we
get $N-2$ spurious solutions, and two solutions for $\lambda$ 
which correspond to genuine
eigenstates. One of these is
a lowering of the two down spin state given at the end of
sec.\ \ref{foursite}; both remain eigenstates for all $\Delta$.
Thus, for eight sites, we get for $\phi=1$
\begin{equation}
\ket{\psi_1} = \ket{11010000} -\ket{10110000} -\ket{11001000} 
+\ket{10011000} + \mbox{translations with $\phi=1$}
\end{equation}
and for $\phi=-1$
\begin{equation}
\ket{\psi_2} = \ket{11010000} + \ket{10110000} + \ket{11001000}
-\ket{10011000} +\mbox{translations with $\phi=-1$}.
\end{equation}

\end{document}